\documentclass[prb,twocolumn]{revtex4}
\usepackage{graphicx}

\begin{document}
\title{Quantum Circuits for Measuring Levin-Wen Operators}
  
\author{N.~E. Bonesteel$^1$ and D.~P. DiVincenzo$^{2,3,4}$}
\affiliation{$^1$Department of Physics and NHMFL, Florida State University, Tallahassee, Florida 32310, USA\\
$^2$Institute for Quantum Information, RWTH Aachen University, D-52056 Aachen,
Germany\\
$^3$Peter Gr\"unberg Institute II, Theoretical Nanoelectronics, Forschungzentrum J\"ulich, J\"ulich, Germany\\
$^4$J\"ulich-Aachen Research Alliance (JARA), Fundamentals of Future Information Technologies}


\begin{abstract}
We construct quantum circuits for measuring the commuting set of vertex and plaquette operators that appear in the Levin-Wen model for doubled Fibonacci anyons.  Such measurements can be viewed as syndrome measurements for the quantum error-correcting code defined by the ground states of this model (the Fibonacci code).  We quantify the complexity of these circuits with gate counts using different universal gate sets and find these measurements become significantly easier to perform if $n$-qubit Toffoli gates with $n =3,4$ and 5 can be carried out directly. In addition to measurement circuits, we construct simplified quantum circuits requiring only a few qubits that can be used to verify that certain self-consistency conditions, including the pentagon equation, are satisfied by the Fibonacci code. 
\end{abstract}

\pacs{03.67.Lx, 03.65.Vf, 05.30.Pr}

\maketitle

\section{Introduction}

The ground states of certain two-dimensional lattice Hamiltonians of a type first introduced by Kitaev\cite{kitaev03} can be used as quantum error-correcting codes known as surface codes.  Quantum information can be stored and protected using these codes when they are defined on lattices with holes (defects).\cite{bravyi98} Fault-tolerant gates can then be carried out either transversally or by deforming the code in order to braid these defects while staying entirely within the code subspace.\cite{raussendorf07a,raussendorf07b,fowler09}
One downside to using the Kitaev surface codes, for which defects behave as Abelian anyons, is that to realize a universal set of fault-tolerant gates at least one gate using a resource costly ``magic state'' distillation process\cite{bravyi05} is required. The same is true for fault-tolerant quantum computation using the so-called color codes.\cite{bombin06,bombin07,bombin11,fowler11,landahl11}  Nevertheless, quantum computation using these surface codes has a number of appealing features, notably the need for only nearest-neighbor gates between qubits in a two-dimensional array and high error thresholds, e.g. $\sim 1\%$ for the Kitaev surface code.\cite{raussendorf07a,raussendorf07b,fowler09,wang11}

Recently K\"onig, Kuperberg, and Reichardt\cite{koenig10} (KKR) outlined a method for fault-tolerant quantum computation using {\it non-Abelian} surface codes.  These codes, which are defined mathematically in terms of the Turaev-Viro topological invariants for 3-manifolds,\cite{turaev92} can be viewed physically as ground states of Levin-Wen models,\cite{levin05}  two-dimensional lattice models which generalize the Kitaev model.\cite{connection}  These models can be used to realize so-called ``doubled'' versions of any consistent anyon theory, including theories of non-Abelian anyons for which braiding is universal for quantum computation. The simplest such universal anyons are the Fibonacci anyons. Here we refer to the corresponding Levin-Wen model as the Fibonacci Levin-Wen model and, following KKR,\cite{koenig10} refer to the ground states of this model as the Fibonacci code.  As shown in Ref.~\onlinecite{koenig10}, when using the Fibonacci code, Fibonacci anyons can be associated with holes in the lattice subject to certain boundary conditions and proper initialization. These Fibonacci anyons can then be used to encode logical qubits and universal quantum computation can be carried out purely by braiding them,\cite{freedman02,bonesteel05,hormozi07} without the need for magic state distillation.   

The Levin-Wen models are defined by a set of commuting vertex and plaquette projection operators which act on qubits (more generally, qudits) associated with the edges of a two-dimensional trivalent lattice. When using the ground states of these models as quantum codes it will be necessary to continually measure these vertex and plaquette operators in order to check for errors, which would then have to be corrected without disturbing the quantum information stored in the topological degrees of freedom of the code.   For the Kitaev surface code, quantum circuits which can be used to measure these operators are known and are fairly straightforward.\cite{dennis02}  For either an $n$-sided plaquette, or a vertex where $n$ edges meet, these measurement circuits each require a single initialized syndrome qubit which is measured after carrying out $n$ controlled-NOT (CNOT) gates.

\begin{figure*}[t]
\begin{center}
\includegraphics[scale=.32]{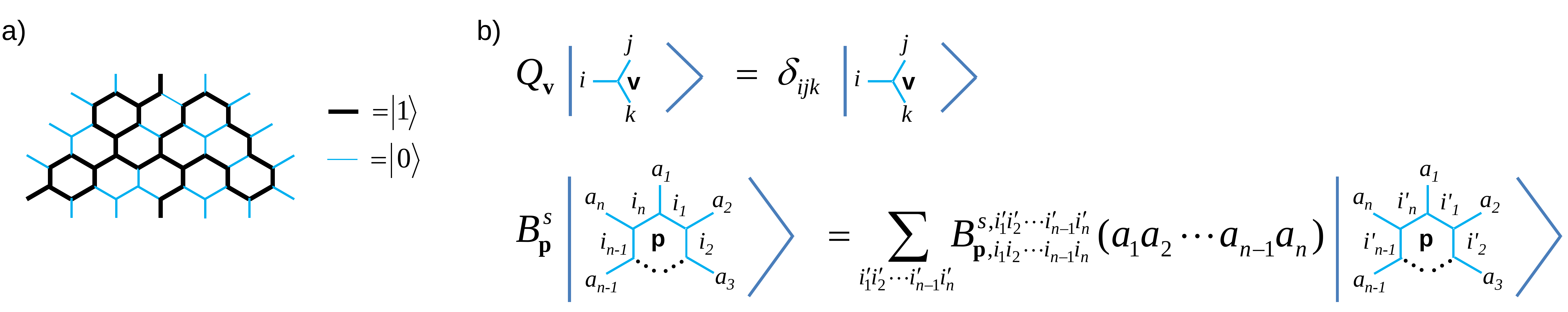}
\end{center}
\caption{(Color online) (a) Example of a trivalent lattice (in this case a honeycomb lattice) on which the Levin-Wen model can be defined.   For the Fibonacci Levin-Wen model a qubit is associated with each edge.  A particular state which satisfies the vertex constraint $Q_{\bf v} = 1$ on each vertex for this model is shown.  Thick edges indicate qubits in the state $|1\rangle$, thin edges indicate qubits in the state $|0\rangle$.  (b) Action of the vertex operator $Q_{\bf v}$ on the three qubits on the edges connected to a trivalent vertex, and of the plaquette operators $B^s_{\bf p}$ on the $2n$ qubits associated with an $n$-sided plaquette.} \label{lattice}
\end{figure*}

The simplicity of the quantum circuits used to measure the vertex and plaquette operators for the Kitaev surface code reflects the Abelian nature of this code. It is natural to ask how complex the quantum circuits need to be to measure the vertex and plaquette operators for the non-Abelian Fibonacci code. In this paper we present explicit quantum circuits for performing such measurements.  These circuits are built in part out of smaller circuits which carry out unitary transformations which have been described both in KKR\cite{koenig10} and, in the context of entanglement renormalization, in Ref.~\onlinecite{koenig09}. Our goal here is to explicitly construct these circuits in terms of standard elements (Toffoli gates, CNOT gates and single-qubit rotations) in an attempt to quantify their complexity.  

The purpose of this work is not to argue that non-Abelian surface codes are viable competitors to the Kitaev surface code.  Indeed, we share the view of many in the field that quantum computation using the Kitaev surface code, given its clear advantages over other fault-tolerant quantum computation schemes, may well provide the best practical route to building a functioning quantum computer.\cite{fowler09,divincenzo09} Here our goal is the more modest one of making a first pass at determining the complexity of syndrome extraction for the significantly less well understood Fibonacci code, which we believe is of intrinsic interest in its own right.  An additional goal of this work is to begin developing a ``dictionary" for translating the mathematical structures which appear in general anyon theories into interesting quantum circuits, some of which we find require only a few qubits and might feasibly be carried out experimentally in the near future.

\section{Levin-Wen Models and the Fibonacci Code}

The Levin-Wen models\cite{levin05} are defined on two-dimensional trivalent lattices such as the hexagonal lattice shown in Fig.~\ref{lattice}(a).  The degrees of freedom of the models are associated with lattice edges which can take on a finite number of labels.  These labels can, in general, be oriented, meaning for each label $i$ there is a dual label $i^*$. If $i=i^*$ then the edge is unoriented.  For the Fibonacci Levin-Wen model there are only two labels $0$ and $1$ and the edges are unoriented ($0=0^*$, $1 = 1^*$).  Thus, for this model, as for the Kitaev surface code, we simply associate a qubit with each edge of the lattice.  The two states of each qubit $|0\rangle$ and $|1\rangle$ then correspond to the two labels 0 and 1, respectively.

For a given trivalent lattice the Levin-Wen Hamiltonian has the form
\begin{eqnarray}
H = - \sum_{\bf v} Q_{\bf v} - \sum_{\bf p} B_{\bf p}.
\end{eqnarray}
Here $Q_{\bf v}$ and $B_{\bf p}$ are projection operators associated with the vertices (labeled ${\bf v}$) and plaquettes (labeled ${\bf p}$) of the lattice.  

The vertex operator $Q_{\bf v}$ acts on the three qubits associated with the edges connected to vertex ${\bf v}$ and is diagonal in the standard $\{|0\rangle$, $|1\rangle\}$ basis. (Here we focus on the Fibonacci Levin-Wen model and so only consider the case when a single qubit is assigned to each edge.)  If these qubits are in the states $|i\rangle$, $|j\rangle$ and $|k\rangle$ the result of applying $Q_{\bf v}$ is determined by the tensor $\delta_{ijk}$ (see Fig.~\ref{lattice}(b)) which, for the Fibonacci Levin-Wen model, is given by,
\begin{eqnarray}
\delta_{ijk} = \left\{\begin{array}{cl} 
1 & {\rm if\ } ijk = 000,011,101,110,111\\
0 & {\rm otherwise.}
\end{array}\right.
\end{eqnarray}

The plaquette operator $B_{\bf p}$ is significantly more complex than $Q_{\bf v}$.  For example, for a hexagonal plaquette, $B_{\bf p}$ acts on the six qubits on the edges of plaquette ${\bf p}$ in a way determined by the state of the six qubits on the edges connected to the plaquette.  $B_{\bf p}$ is therefore a twelve-qubit interaction (in general a $2n$-qubit interaction for an $n$-sided plaquette).  For the Fibonacci Levin-Wen model the precise form of the plaquette projection operator is,
\begin{eqnarray}
B_{\bf p} = \frac{1}{1+\phi^2}
\left(B_{\bf p}^0 + \phi B_{\bf p}^1\right),
\label{Eq:Bp}
\end{eqnarray}
where $B_{\bf p}^s$ for $s=0$ and 1 are plaquette operators associated with the label $s$ and $\phi = (\sqrt{5}+1)/2$ is the golden ratio. The action of $B_{\bf p}^s$ on an $n$-sided plaquette is shown in Fig.~\ref{lattice}(b) where,
\begin{eqnarray} 
&&B^{s,i^\prime_1 i^\prime_2 \cdots i^\prime_{n-1}i^\prime_n}_{{\bf p},i_1 i_2 \cdots i_{n-1} i_n}(a_1 a_2 \cdots a_{n-1} a_n)\\ 
&&~~= 
F^{a_1i_n i_1}_{si_1^\prime i_n^\prime}
F^{a_2i_1 i_2}_{si_2^\prime i_1^\prime}
\cdots
F^{a_{n-1} i_{n-2} i_{n-1}}_{si_{n-1}^\prime i_{n-2}^\prime}
F^{a_n i_{n-1} i_n}_{s i_n^\prime i_{n-1}^\prime}. \label{Eq:Bps}
\nonumber
\end{eqnarray}
Here the six-indexed tensor $F^{ijk}_{lmn}$, along with $\delta_{ijk}$, forms the basic data of a so-called tensor category --- the mathematical framework for a general anyon theory, in this case the theory of Fibonacci anyons.  The $F$ and $\delta$ tensors satisfy certain self-consistency conditions which, among other things, guarantee that the operators $B_{\bf p}^s$ and $Q_{\bf v}$ all commute with each other.\cite{levin05,bp0}  Note that since the Fibonacci Levin-Wen model is unoriented, in (\ref{Eq:Bps}) we have assumed $i= i^*$ for all labels.  The precise form of the $F$ tensor for this model is given in Sec.~\ref{fmove_sec}.   

When using the ground states of the Levin-Wen model as quantum error-correcting codes the commuting vertex and plaquette projection operators $Q_{\bf v}$ and $B_{\bf p}$ should be viewed as stabilizers.  The code space is then defined by the requirement that $Q_{\bf v} = 1$ on each vertex and $B_{\bf p} = 1$ on each plaquette.  For the Fibonacci code the constraint $Q_{\bf v} = 1$ projects the Hilbert space onto the space spanned by states in which edges in the state $|1\rangle$ form branching loop configurations (see Fig.~\ref{lattice}(a)), while the plaquette constraint $B_{\bf p} = 1$ leads to particular quantum superpositions of these states. As described in KKR,\cite{koenig10} when these code states are defined on lattices with holes that have certain boundary conditions on their edges, these holes (or defects) can realize a ``doubled'' version of the anyon theory characterized by the $F$ and $\delta$ tensors.  For the Fibonacci code this means that these defects can encode two types of Fibonacci anyons with opposite chiralities.  As further shown in KKR,\cite{koenig10} with proper initialization these defects can be forced to encode Fibonacci anyons of a particular chirality.  These chiral anyons can then be used to encode qubits and braided in order to carry out universal quantum computation.

In this paper we focus on the problem of how to measure the stabilizers $Q_{\bf v}$ and $B_{\bf p}$ for the Fibonacci code.  In the passive approach to the Levin-Wen model envisioned in KKR,\cite{koenig10} rather than engineering the Levin-Wen Hamiltonian to realize the Fibonacci code it will be necessary to continually measure these operators in order to detect errors which can then be corrected.  

\section{Quantum Circuit to Measure $Q_{\bf v}$}
\label{Sec:Qv}

\begin{figure}[t]
\centerline{
\includegraphics[scale=.4]{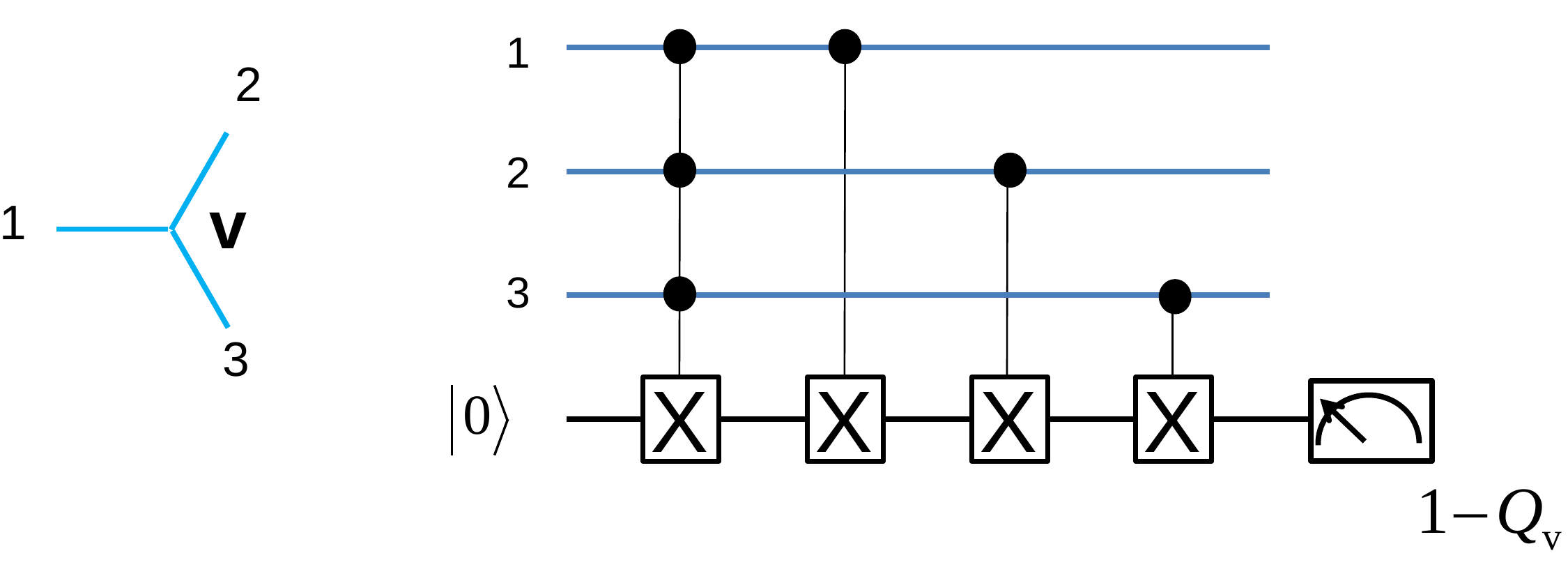}
}
\ 
\caption{(Color online) Quantum circuit which can be used to measure $Q_{\bf v}$ for the Fibonacci code.} \label{Qv}
\end{figure}

The measurement of $Q_{\bf v}$ for the Fibonacci code is straightforward and not significantly more difficult to carry out than the analogous measurement for the Kitaev surface code. A quantum circuit which carries out a quantum non-demolition measurement of $Q_{\bf v}$ is shown in Fig.~\ref{Qv}.  The circuit acts on the three qubits associated with a given vertex as well as a fourth syndrome qubit initialized in the state $|0\rangle$.  After carrying out the circuit the syndrome qubit is measured. If it is found to be in the state $|0\rangle$ then $Q_{\bf v} = 1$ for this vertex and the vertex constraint is satisfied, if not then $Q_{\bf v} = 0$ and the vertex constraint is violated.  

The most difficult part of the $Q_{\bf v}$ circuit to carry out is likely to be the four-qubit Toffoli gate which performs a NOT gate on the syndrome qubit if and only if the state of each of the three vertex qubits is $|1\rangle$. (Here and throughout it should be understood that an $n$-qubit Toffoli gate is a gate with $n-1$ control qubits and one target qubit.) This four-qubit Toffoli gate is the first of several $n$-qubit Toffoli gates required in our constructions, all of which are directly related to the non-Abelian nature of the Fibonacci code.  Here this gate is needed to allow for the loop branching associated with the fact that $\delta_{111} = 1$. 

In what follows we will be interested in quantifying the complexity of the quantum circuits we construct.  Of course the notion of quantum circuit complexity is somewhat ill-defined and depends, among other things, on what we take as our primitive gate set.  This in turn will depend on the particular hardware of the quantum computer being considered.  

Accurate three-qubit Toffoli-class gates have recently been been carried out experimentally using superconducting qubits\cite{mariantoni11,fedorov12,reed12} and trapped ions.\cite{monz09} Motivated by this, we take one primitive gate set to consist of three-qubit Toffoli gates, CNOT gates and single-qubit rotations. An $n$-qubit Toffoli gate can then be carried out using $4n-12$ three-qubit Toffoli gates if $n-3$ additional qubits are available.\cite{barenco95_1} These additional qubits need not be initialized and their states are left unchanged once the full $n$-qubit Toffoli gate is carried out. Thus nearby code qubits which are not being acted on directly by the operator under measurement can be used. With this construction we can count the total number of three-qubit Toffoli gates (or, simply, Toffoli gates), CNOT gates and single-qubit rotations required to carry out a given circuit.  For the case of the four-qubit Toffoli gate appearing in our $Q_{\bf v}$ circuit this count gives $4$ Toffoli gates.  The total gate count for our $Q_{\bf v}$ circuit is then 4 Toffoli gates and 3 CNOT gates.  This can be contrasted with the analogous circuit for the Kitaev surface code which, when acting on a trivalent vertex, would require only 3 CNOT gates (it is, in fact, identical to the circuit shown in Fig.~\ref{Qv} with the four-qubit Toffoli gate removed).\cite{dennis02}

For a second gate count we assume that the $n$-qubit Toffoli gates which appear in our circuits are themselves primitive gates.  By this count, our $Q_{\bf v}$ circuit consists of 1 four-qubit Toffoli gate and 3 CNOT gates.  We note that there are proposals for carrying out single-step $n$-qubit Toffoli-class gates using trapped ions,\cite{cirac95} superconducting qubits,\cite{lin06} and neutral atoms interacting with cavity photons;\cite{duan05} in addition, it has been observed that these gates are efficiently achieved if one of the qubits has $n$ available quantum levels.\cite{ralph07} Of course $n$-qubit Toffoli gates can also be simulated using the usual primitive gate set consisting of CNOT gates and single-qubit rotations.\cite{toffoli}   However, as we have seen with our $Q_{\bf v}$ measurement circuit, and as will become more clear in what follows, the ability to directly carry out accurate $n$-qubit Toffoli gates (with $n = 3,4$ and 5) will give a strong advantage when carrying out quantum computation using the Fibonacci code.

\begin{figure}[t]
\begin{center}
\includegraphics[scale=.35]{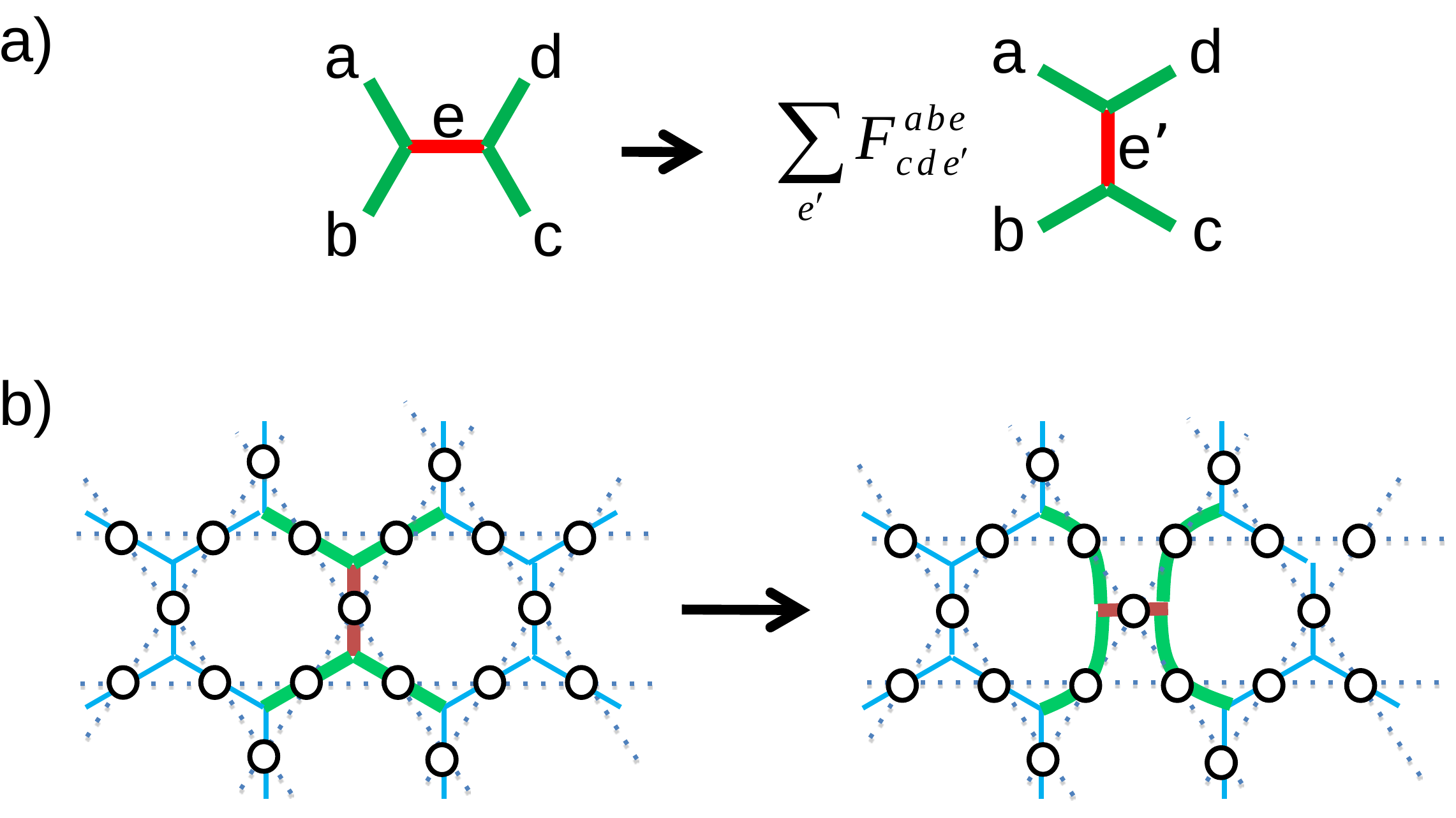}
\end{center}\caption{(Color online) (a) An $F$-move, a five-qubit unitary operation defined in terms of the tensor $F^{abe}_{cde^\prime}$.  (b) Action of an $F$-move on the abstract trivalent lattice of the Fibonacci code which illustrates the decoupling of this lattice from the physical qubits.  In this example the qubits (open circles) are arranged in a Kagome lattice and lie on the edges of an initial trivalent (hexagonal) lattice.  After the $F$-move the edges of the new trivalent lattice must be distorted if they are forced to coincide with the physical qubit lattice.} \label{f_lattice}
\end{figure}

\begin{figure*}
\begin{center}
\includegraphics[scale=.35]{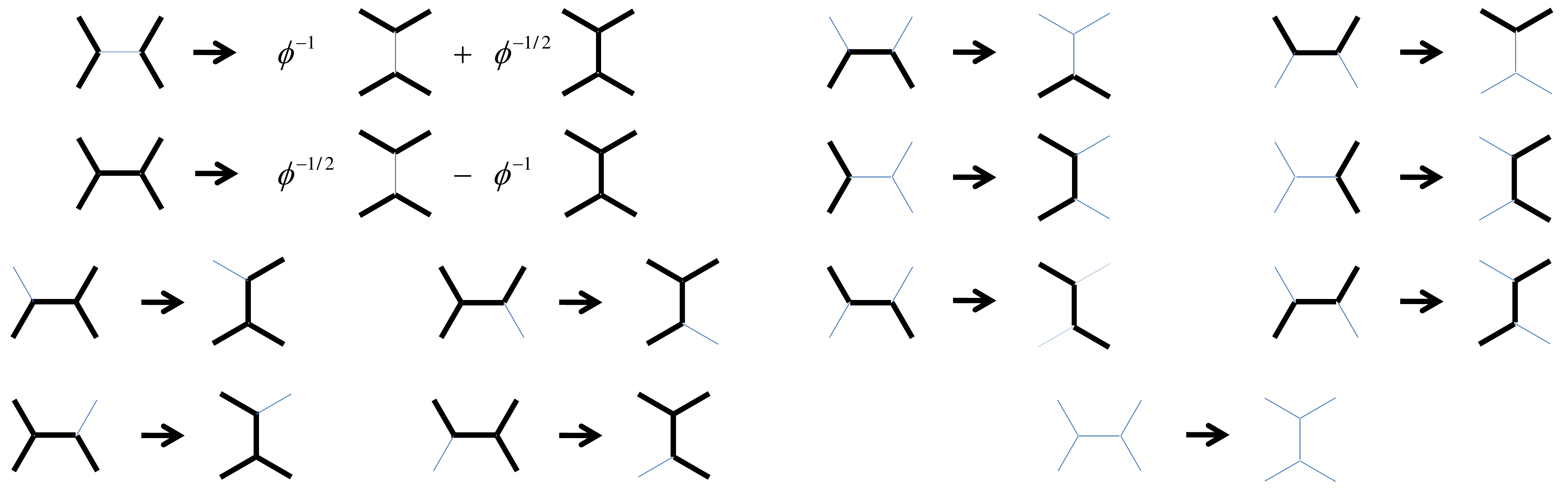}
\end{center}
\caption{(Color online) $F$-move for Fibonacci anyons.  Under this $F$-move a unitary transformation is performed on the qubit associated with the edge which goes from horizontal to vertical conditioned on the state of the qubits on the other four edges.  As in Fig.~\ref{lattice} thick lines indicate edges in the state $|1\rangle$ and thin lines indicate edges in the state $|0\rangle$. Only those states which satisfy the $Q_{\bf v}= 1$ constraint are shown.} \label{f_move}
\end{figure*}

Despite requiring a four-qubit Toffoli gate, the $Q_{\bf v}$ measurement circuit shown in Fig.~\ref{Qv} is relatively simple, reflecting the simplicity of the vertex operator. In what follows we turn to the more difficult problem of measuring the plaquette operator $B_{\bf p}$. For this case a brute force approach to constructing a circuit which measures the appropriate operator acting on the edges of a plaquette for each possible state of the edges connected to that plaquette is problematic. Fortunately, there is a useful resource which simplifies the problem greatly --- the $F$-move.

\section{$F$-Move}
\label{fmove_sec}

When using the Fibonacci code, the physical qubits of a quantum computer may be fixed in space and may even form a rigid lattice.  However, this physical lattice need not be the same as that formed by the edges of the abstract trivalent lattice used to define the code.  Indeed, as emphasized in KKR,\cite{koenig10} this abstract trivalent lattice should be thought of as fluid and constantly changing throughout the computation.  These changes are accomplished by carrying out $F$-moves, processes which locally redraw the trivalent lattice while reassigning the physical qubits to new lattice edges and carrying out an appropriate unitary operation.

Specifically, when carrying out an $F$-move five edges of the lattice are redrawn as shown in Fig.~\ref{f_lattice}(a) while a unitary transformation determined by the six indexed tensor $F^{abe}_{cde^\prime}$ (the same $F$ tensor which appears in (\ref{Eq:Bps})) is applied to the five qubits associated with these edges.  This five-qubit unitary is a controlled operation on the qubit labeled $e$ in Fig.~\ref{f_lattice}(a) contingent on the states of the other four qubits (labeled $abcd$). The usefulness of the $F$-move here derives from the fact that if one starts in a ground state of a given Levin-Wen model on a particular trivalent lattice then, after performing an $F$-move, the resulting state will be a ground state of the new Levin-Wen model defined on the new trivalent lattice.\cite{koenig09} This is true even though this lattice has decoupled from the physical qubits, as illustrated in Fig.~\ref{f_lattice}(b).

It was shown in KKR\cite{koenig10} that the ability to decouple the abstract trivalent lattice from the physical qubits with $F$-moves is an important resource for carrying out quantum computation using the Fibonacci code.  For example, by carrying out sequences of $F$-moves one can deform the code to perform Dehn twists on the trivalent lattice which can then be used to braid defects encoding Fibonacci anyons.\cite{koenig10} Since the braiding of Fibonacci anyons is universal for quantum computation, this means that one can perform a universal set of gates while staying inside the Fibonacci code subspace without the need for magic state distillation.

The $F$-move for the Fibonacci code is represented graphically in Fig.~\ref{f_move}.  This figure, together with Fig.~\ref{f_lattice}(a), can serve as a definition of the $F$ tensor for Fibonacci anyons.  The effect of carrying out an $F$-move is only shown for those states which satisfy the vertex constraint (i.e. for which $Q_{\bf v}$ = 1 for all vertices). When defining the Levin-Wen models, the $F$ tensor is assumed to vanish when acting on those states which violate the vertex constraint.\cite{levin05}  Here we will assume before applying any $F$-move that it has been verified that $Q_{\bf v} = 1$ on each relevant vertex of the initial trivalent lattice.  The structure of the $F$-move then guarantees that the vertex constraint will continue to be satisfied on the new trivalent lattice.

\begin{figure}[b]
\begin{center}
\includegraphics[scale=.4]{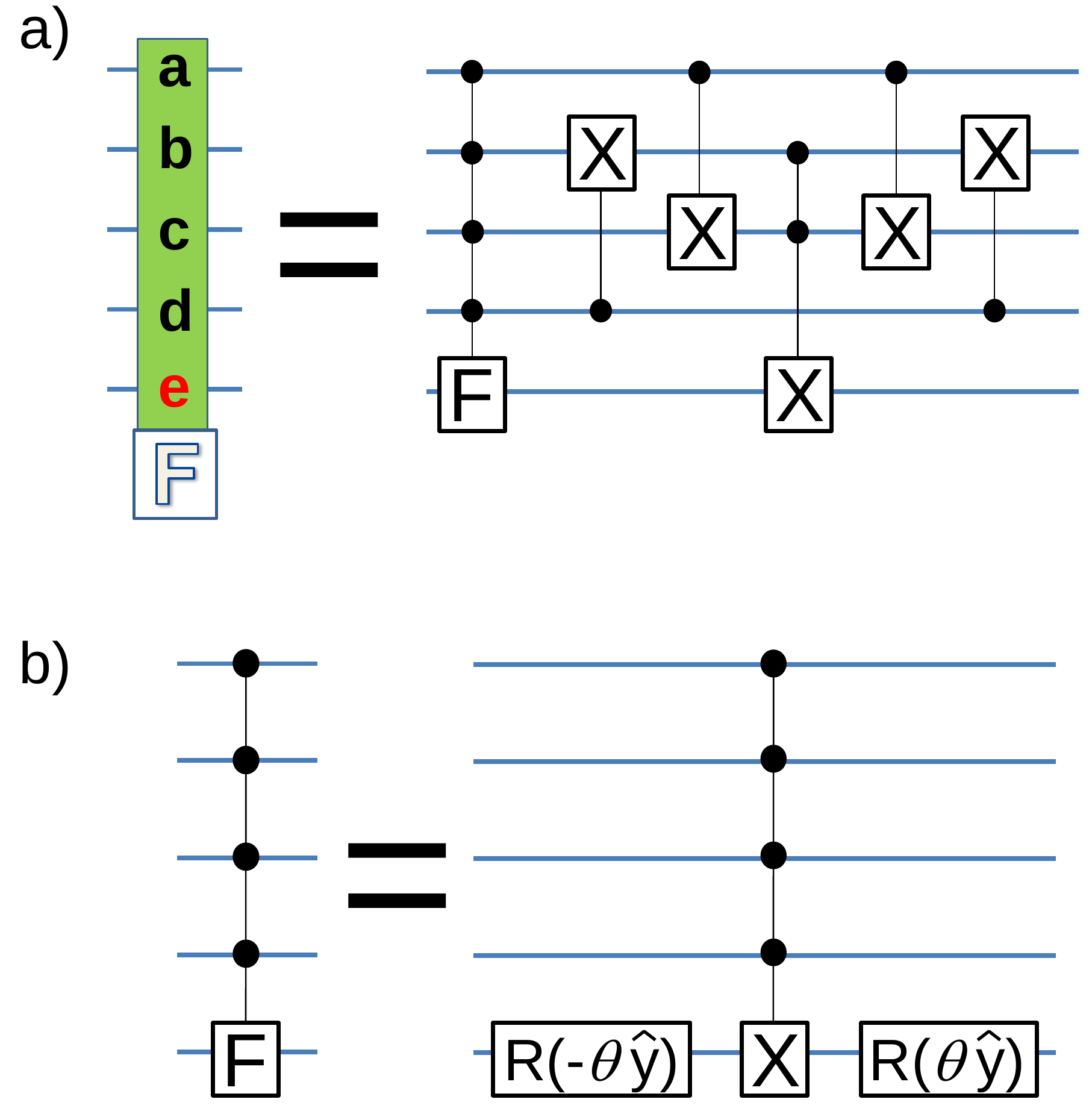}
\end{center}\caption{(Color online) (a) Quantum circuit which carries out an $F$-move for the Fibonacci code (the $2\times 2$ matrix $F$ is given in Eq.~\ref{fmatrix}).  The labels $abcde$ refer to the same labels in Fig.~\ref{f_lattice}(a). (b) Five-qubit controlled-$F$ gate expressed in terms of a five-qubit Toffoli gate. Here $R(\pm \theta\hat y) = e^{\pm i\theta \sigma_y/2}$ are single-qubit rotations about the $y$ axis with $\theta = \tan^{-1} \phi^{-1/2}$ for which $R(\theta \hat y)XR(-\theta \hat y) = F$.} \label{f}
\end{figure}

A quantum circuit which acts on five qubits at a time and which carries out the $F$-move defined in Fig.~\ref{f_move} for those states satisfying the vertex constraint is shown in Fig.~\ref{f}.  In this figure the labels $abcde$ refer to the same labels shown in Fig.~\ref{f_lattice}(a).  Although it is not immediately apparent from its structure, one can readily check that this circuit has the symmetries of the $F$ tensor\cite{levin05} (e.g., $F^{abe}_{cde^\prime} = F^{cde}_{abe^\prime} = F^{bae}_{dce^\prime}$).  Note also that the circuit squares to 1 (since $F^2 = 1$, see below), so the same circuit can be used for the inverse transformation.  As described above, this $F$ circuit carries out a particular operation on the qubit labeled $e$ depending on the state of the other four qubits labeled $abcd$ which are themselves left unchanged at the end of the circuit. The $F$ circuit can therefore be viewed as a generalized Toffoli-class gate. Because the four control qubits are not equivalent, it is important to label these qubits in our $F$ circuit as we have done in the green box in Fig.~\ref{f_move}. This notation will be useful when we embed the $F$ circuit into larger circuits acting on more than five qubits.  

\begin{figure*}[t]
\begin{center}
\includegraphics[scale=.35]{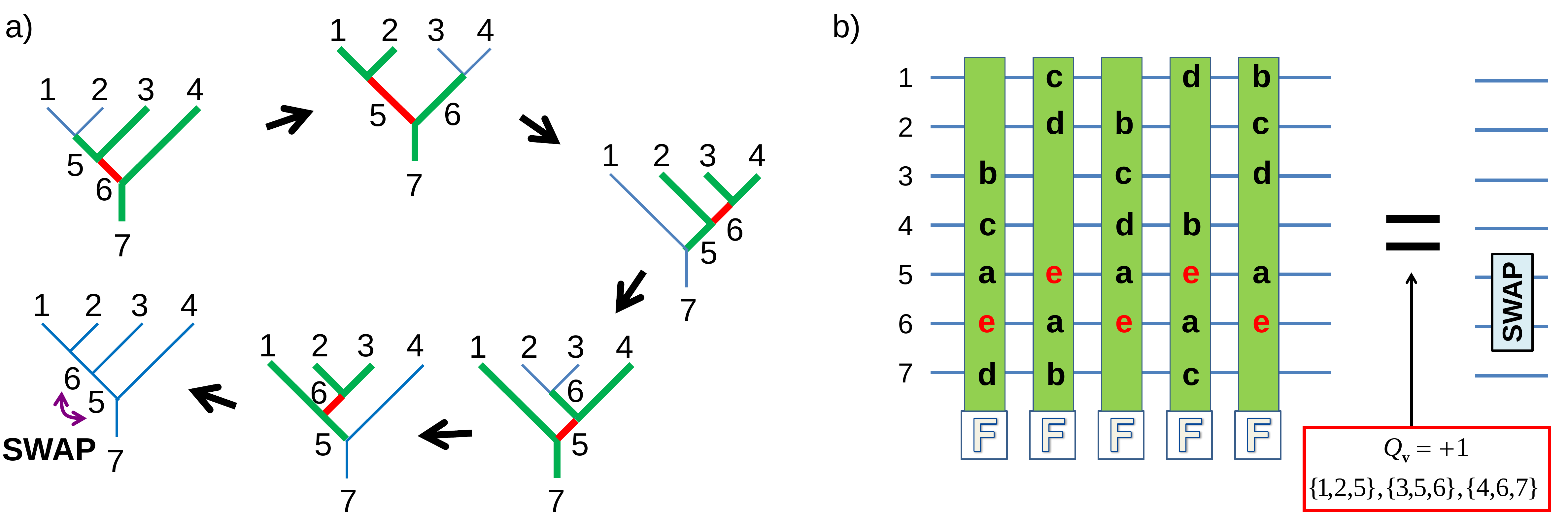}
\end{center}
\caption{(Color online) (a) The pentagon equation, a self-consistency condition which the $F$-move must satisfy.  As shown here, the pentagon equation corresponds to a series of $F$-moves which take a particular 7 edged lattice (upper left) back to an identical lattice (lower left) while two of the qubits associated with the lattice edges are swapped.  Here and in subsequent figures the edges associated with the initial state before each $F$-move are color coded as in Fig.~\ref{f_lattice}. (b) The pentagon equation as a quantum circuit identity.  The sequence of $F$-moves shown in (a) are carried out by repeatedly applying the $F$ circuit defined in Fig.~\ref{f}.  The labels $abcde$ in each green box refer to the labels in Fig.~\ref{f}.  The circuit equality holds provided the vertex constraint $Q_{\bf v} = 1$ is satisfied on all three vertices in the initial lattice.  In the figure, the triplets of numbers given below ``$Q_{\bf v} = 1$'' in the red box indicate the qubits which meet at these vertices.} \label{pentagon}
\end{figure*}

\begin{figure}[b]
\centerline{\includegraphics[scale=.45]{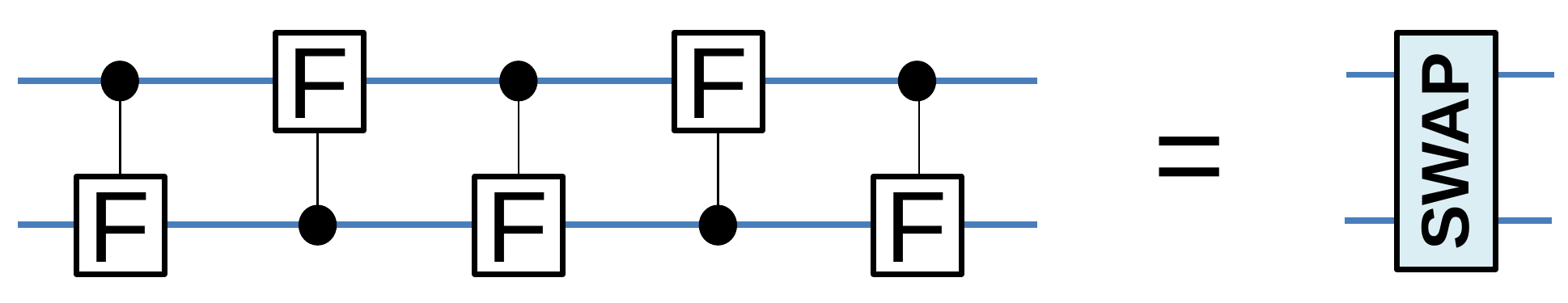}}
\caption{(Color online) Simple two-qubit circuit identity obtained by setting the five effective control qubits (qubits 1,2,3,4, and 7) in the pentagon circuit identity shown in Fig.~\ref{pentagon}(b) to the state $|1\rangle$.} \label{pentagon_simple}
\end{figure}

At the heart of the $F$ circuit is the five-qubit controlled-$F$ gate where $F$ is the $2\times 2$ unitary matrix acting on qubit $e$ when $a=b=c=d=1$,
\begin{eqnarray}
F = \left(\begin{array}{cc} \phi^{-1} & \phi^{-1/2} \\
\phi^{-1/2} & - \phi^{-1}
\end{array}\right).
\label{fmatrix}
\end{eqnarray}
The remaining Toffoli gate and CNOT gates take care of all other cases for which the outcome is essentially fixed by the vertex constraint. As stated above, this circuit is designed to carry out an $F$-move only on those states which satisfy the $Q_{\bf v}= 1$ constraint on all vertices.  In what follows we will always assume it has been verified that the vertex constraint is satisfied before applying the $F$ circuit.
 
Figure \ref{f}(b) shows how to carry out the five-qubit controlled-$F$ gate using a five-qubit Toffoli gate and two single-qubit rotations.  This simple construction is possible because $F^2 = 1$ and $\det F = -1$. As for the four-qubit Toffoli gate appearing in the measurement circuit for $Q_{\bf v}$, the appearance of this five-qubit Toffoli gate can be traced back to the fact that loops are allowed to branch in the Fibonacci code and is a direct consequence of the non-Abelian nature of this code.  Using the construction of Ref.~\onlinecite{barenco95_1} described above this five-qubit Toffoli gate can be carried out using 8 conventional Toffoli gates. The total gate count for our $F$ circuit is then 9 Toffoli gates, 4 CNOT gates and 2 single-qubit rotations.  Alternatively, if we treat $n$-qubit Toffoli gates as primitives, the gate count is 1 five-qubit Toffoli gate, 1 Toffoli gate, 4 CNOT gates and 2 single-qubit rotations.  Given the importance of carrying out $F$-moves when using the Fibonacci code,\cite{koenig10} the ability to accurately carry out this five-qubit Toffoli gate can be viewed as an important experimental threshold for realizing this type of quantum computation. 

\section{Pentagon Equation}
\label{Sec:pentagon}

The $F$-move satisfies an important self-consistency condition known as the pentagon equation.  The pentagon equation can be represented as a sequence of $F$-moves on a seven-edged trivalent lattice as shown in Fig.~\ref{pentagon}(a).  In a quantum computer, the lattice edges would be associated with qubits, labeled 1 through 7 in Fig.~\ref{pentagon}(a).  As one follows this sequence of $F$-moves, the trivalent lattice is repeatedly redrawn while the qubits, which can be considered fixed in physical space, are reassigned to the new lattice edges after each $F$-move.  By the time one has gone all the way around the pentagon the trivalent lattice has returned to its original form. However, the qubits associated with two of the edges (labeled $5$ and $6$ in the figure) are swapped.   

The process of carrying out this sequence of five $F$-moves and the resulting qubit swap can be translated into the quantum circuit identity shown in Fig.~\ref{pentagon}(b).  We refer to the left-hand side of this identity as the pentagon circuit.  The solid green rectangles in the pentagon circuit represent the five-qubit $F$ circuit shown in Fig.~\ref{f} and the corresponding $abcde$ labels are the same as the labels shown in Fig.~\ref{f}.  Again we assume that before carrying out the pentagon circuit it has been verified that $Q_{\bf v} = 1$ on each of the two vertices of the initial trivalent lattice.  It is only for this case that the circuit identity shown in Fig.~\ref{pentagon}(b) holds (for clarity these vertices are labeled by their associated qubits inside the red box under the equals sign in this figure).

In the pentagon circuit two of the qubits (qubits $5$ and $6$) are acted on while the remaining qubits play the role of control qubits. Simpler quantum circuits can be constructed by fixing these five effective control qubits to be in a particular state.  For example, if we fix all the qubits except for $5$ and $6$ to be in the state $|1\rangle$ then the pentagon circuit reduces to the simple two-qubit circuit shown on the left-hand side of the circuit identity in Fig.~\ref{pentagon_simple}.  This simplified pentagon circuit consists of five controlled-$F$ gates with alternating control qubits, and the net effect of this sequence of gates is a SWAP gate. Note that when qubits 5 and 6 are both in the state $|0\rangle$ and all other qubits are in the state $|1\rangle$ the vertex constraint is violated in the full seven-qubit pentagon circuit. However, in this case the simplified pentagon circuit merely carries out the identity operation, which is consistent with swapping the two qubits.  Therefore the expression shown in Fig.~\ref{pentagon_simple} is an exact circuit identity, regardless of the vertex constraint. 

We note the resemblance of this circuit identity to the familiar three CNOT construction of the SWAP gate.\cite{feynman85,barenco95_2}  In our case, the circuit identity shown in Fig.~\ref{pentagon_simple} represents the nontrivial part of the pentagon equation which uniquely fixes the form of the matrix $F$ (up to an arbitrary and irrelevant phase choice for the off-diagonal matrix elements).  We envision that this circuit identity may be useful for calibrating the $F$ operation.  For example, one can imagine tuning $F$ until it can be verified by quantum process tomography that five controlled-$F$ gates with alternating control qubits indeed produce a SWAP gate.

\begin{figure*}[t]
\begin{center}
\includegraphics[scale=.37]{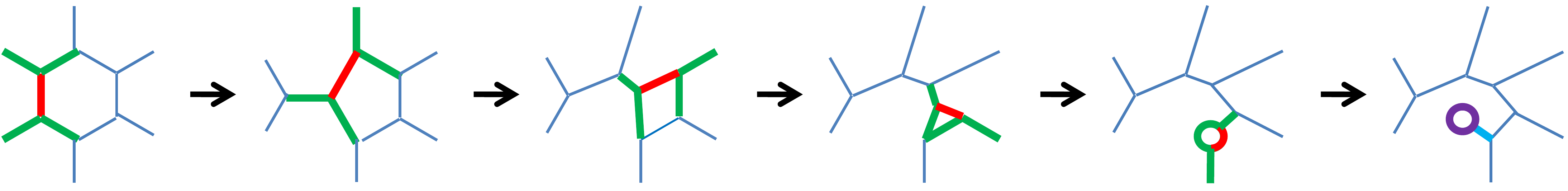}
\end{center}
\caption{(Color online) Reduction of a hexagonal plaquette to a tadpole through a sequence of $F$-moves.} \label{plaquette_reduction}
\end{figure*}

\section{Quantum Circuit to Measure $B_{\bf p}$}

\begin{figure}[b]
\centerline{\includegraphics[scale=.4]{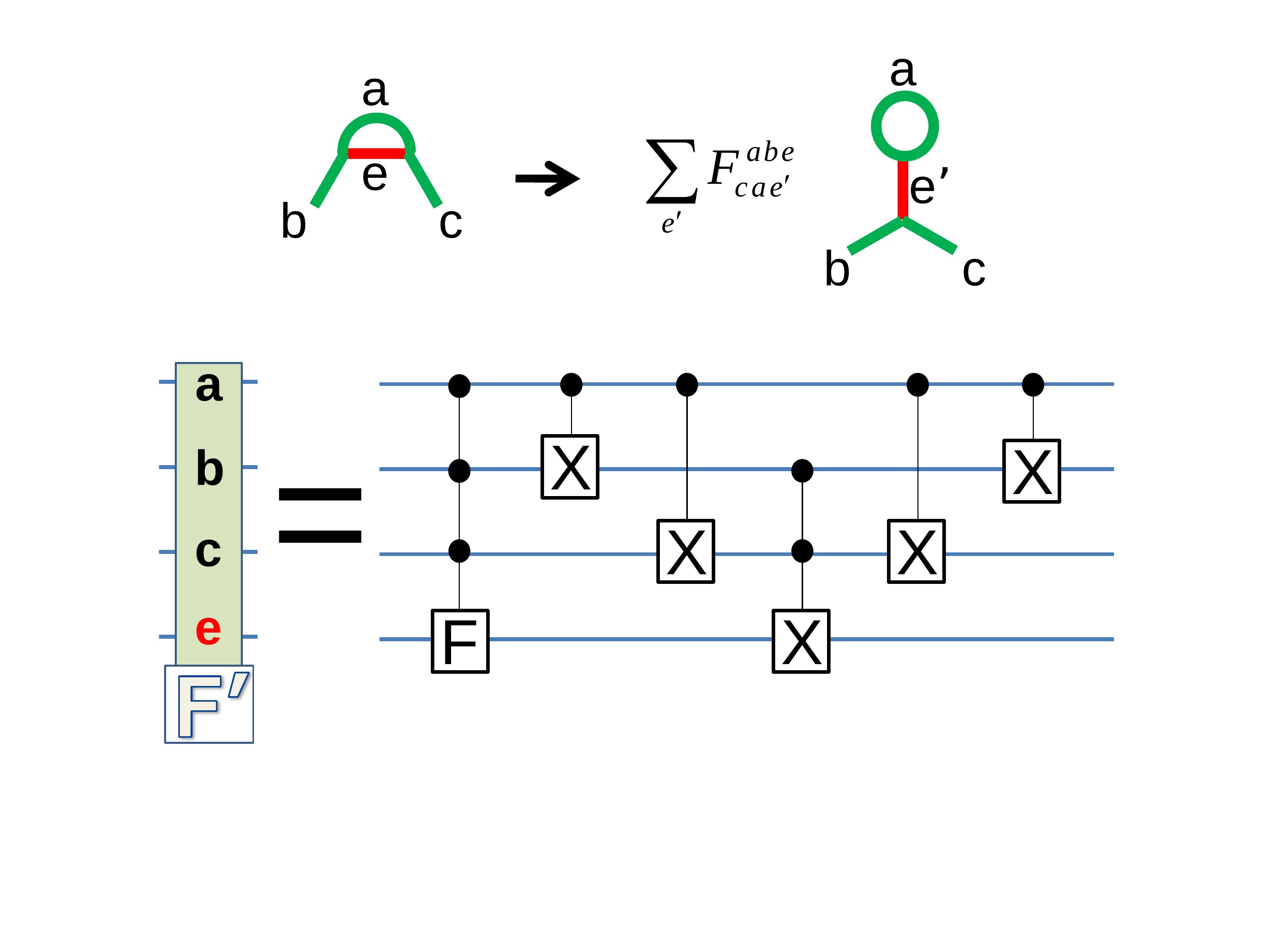}}
\caption{(Color online) Reduced four-qubit $F$-move obtained by identifying the qubits labeled $a$ and $d$ in Fig.~\ref{f}.} \label{fprime}
\end{figure}

We now turn to constructing a quantum circuit to measure the plaquette operator $B_{\bf p}$.  To do this we use a method inspired by the entanglement renormalization scheme of Ref.~\onlinecite{koenig09}. The essential idea is that through a sequence of $F$-moves any $n$-sided plaquette can be reduced to a 1-sided plaquette with a single external line, i.e. a ``tadpole.''  One such sequence of $F$-moves which reduces a hexagonal plaquette to a tadpole is shown in Fig.~\ref{plaquette_reduction}.  Note that the final $F$-move in this sequence acts on four qubits rather than five.  A quantum circuit which carries out this reduced $F$-move, obtained by identifying the qubits labeled $a$ and $d$ in the circuit shown in Fig.~\ref{f}, is shown in Fig.~\ref{fprime}. (Gate counts for this reduced $F$ circuit: 5 Toffoli gates, 4 CNOT gates, and 2 single-qubit rotations, or 1 four-qubit Toffoli gate, 1 Toffoli gate, 4 CNOT gates, and 2 single-qubit rotations.)

It was shown in Ref.~\onlinecite{koenig09} that the plaquette operator $B_{\bf p}$ commutes with $F$-moves, i.e. after each $F$-move shown in Fig.~\ref{plaquette_reduction} the value of $B_{\bf p}$ is unchanged even as the plaquette size is reduced.  This is equivalent to the statement that if we start with a plaquette in a ground state of the Levin-Wen model (meaning $Q_{\bf v} = 1$ on each vertex and $B_{\bf p} = 1$ for the plaquette) then, after each $F$-move, the qubits will continue to be in the ground state of the Levin-Wen model for the new lattice.  Thus, after each $F$-move, it will still be true that $Q_{\bf v} = 1$ on each vertex and $B_{\bf p} = 1$ on the reduced plaquette. This means that after performing the ``disentangling'' reduction of the $n$-sided plaquette to a tadpole one need only measure $B_{\bf p}$ for the tadpole to measure $B_{\bf p}$ for the original plaquette.  Since the tadpole only consists of two qubits this measurement is straightforward.

\begin{figure}[b]
\centerline{\includegraphics[scale=.35]{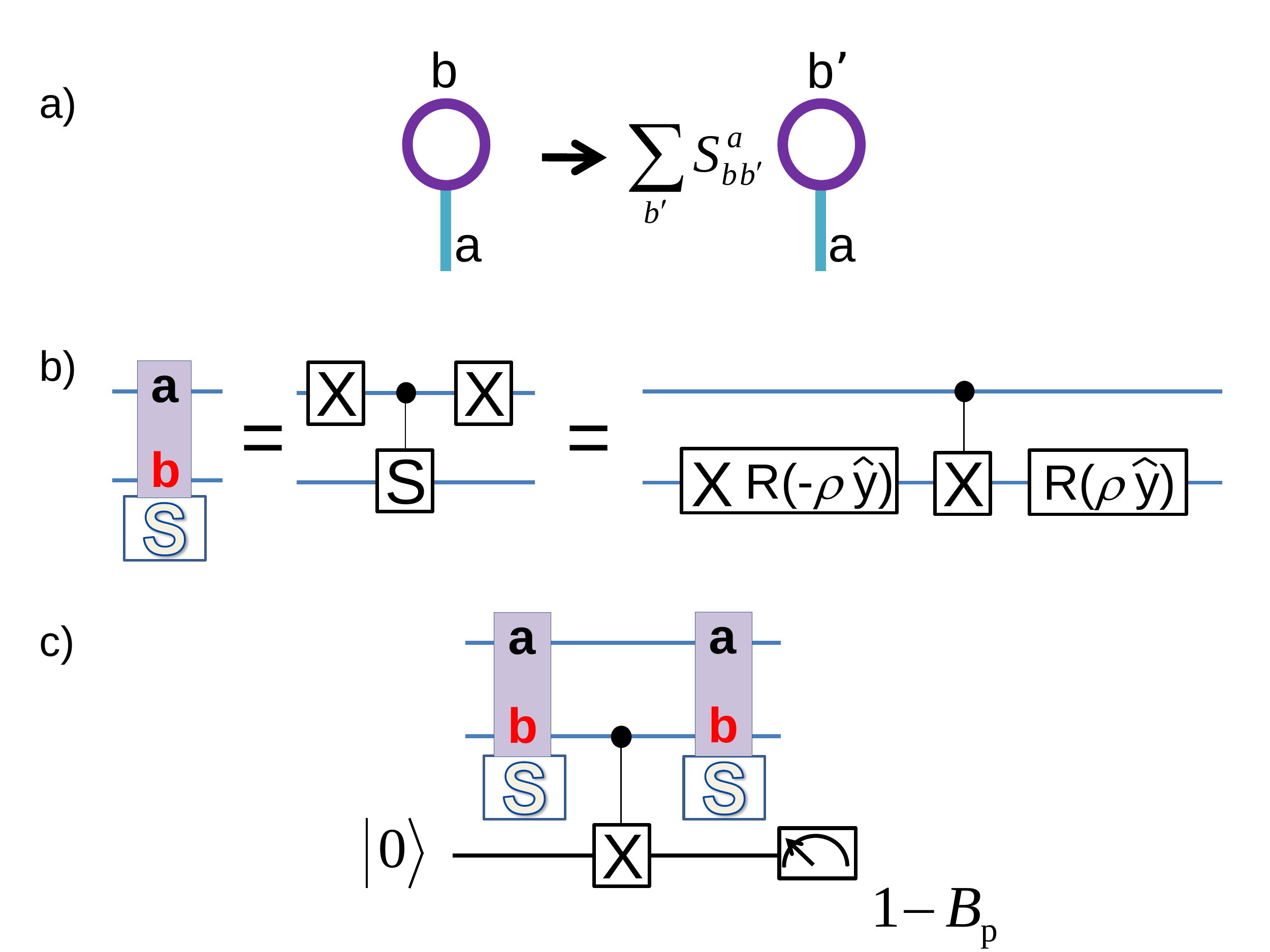}}
\caption{(Color online) (a) $S$ transformation acting on a two-qubit tadpole.  The tensor $S^a_{bb^\prime}$ is defined in the text. (b) $S$ circuit which carries out an $S$ transformation.  The $2\times 2$ matrix $S$ is given in Eq.~\ref{smatrix}.  Here $R(\pm\rho\hat y) = e^{\pm i\rho\sigma_y/2}$ are single-qubit rotations about the $y$ axis with $\rho = \tan^{-1}  \phi^{-1}$ for which $R(\rho \hat y)XR(-\rho \hat y) = S$. (c) Quantum circuit which uses the $S$ circuit to measure $B_{\bf p}$ for a two-qubit tadpole.}\label{s}
\end{figure}

\begin{figure*}[t]
\begin{center}
\includegraphics[scale=.55]{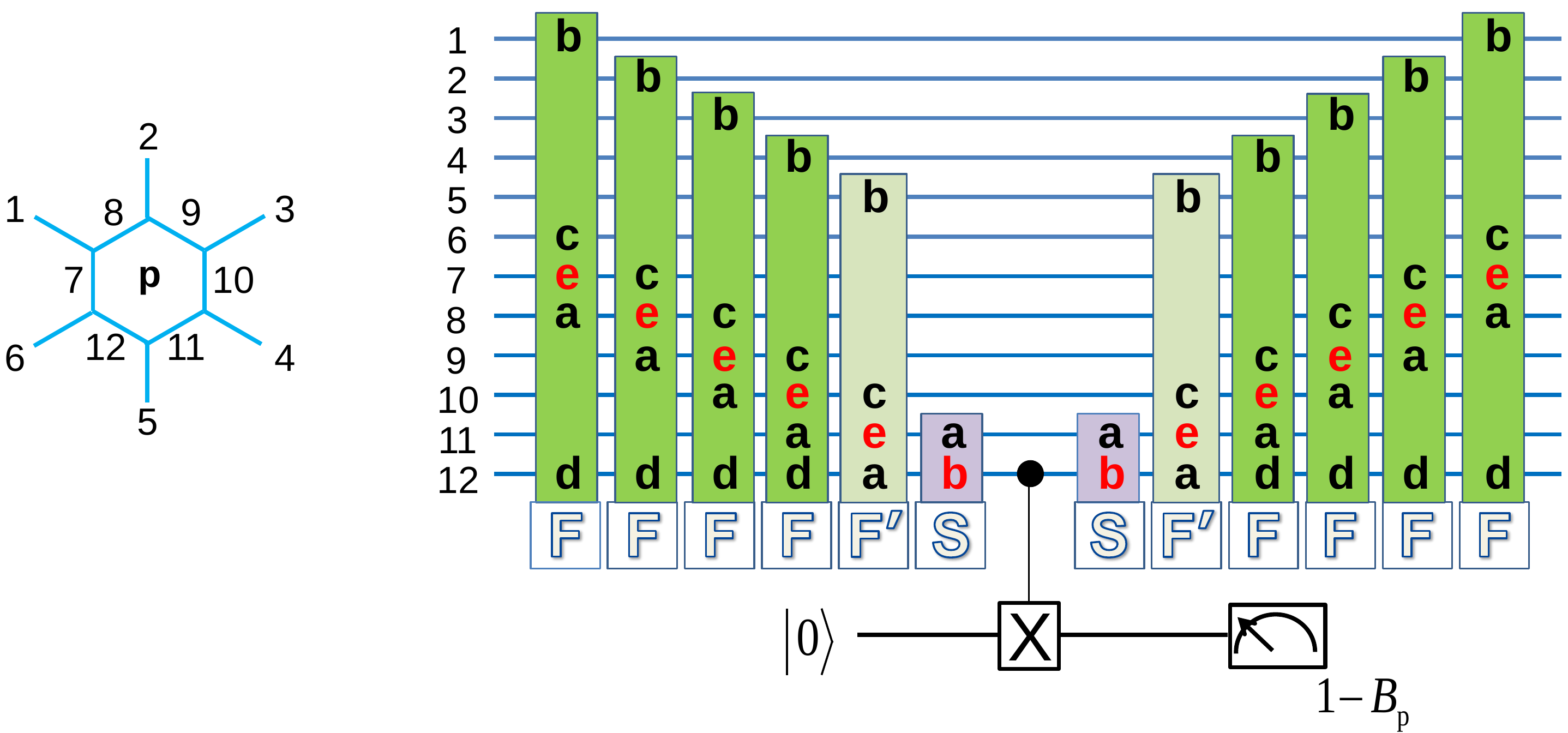} 
\end{center}
\caption{(Color online) Quantum circuit which can be used to measure $B_{\bf p}$ for the Fibonacci code on a hexagonal plaquette based on the plaquette reduction shown in Fig.~\ref{plaquette_reduction}.  It must be verified that $Q_{\rm v} = 1$ on each of the six vertices of the plaquette before carrying out the circuit.} \label{Bp}
\end{figure*}

For a tadpole there is a unique eigenstate of $B_{\bf p}$ with eigenvalue 1,\cite{koenig09}
\begin{eqnarray}
|\psi_{B_{\bf p} = 1}\rangle = |0\rangle(|0\rangle + \phi|1\rangle)/\sqrt{1+\phi^2}.
\end{eqnarray}
Here the first qubit is the external line (tail of the tadpole) and the second qubit is the 1-sided plaquette (head of the tadpole).   The two-dimensional Hilbert space of states orthogonal to $|\psi_{B_{\bf p} = 1}\rangle$ which satisfy the vertex constraint will then have $B_{\bf p} = 0$.  This space is spanned by the states
\begin{eqnarray}
|\psi_{B_{\bf p} = 0},a\rangle &=& |0\rangle(\phi|0\rangle - |1\rangle)/\sqrt{1+\phi^2},\label{bp2}\\
|\psi_{B_{\bf p} = 0},b\rangle &=& |1\rangle |1\rangle.\label{bp3}
\end{eqnarray}

To measure $B_{\bf p}$ for this simple two-qubit system we first rotate the head qubit of the tadpole so that it is in the state $|0\rangle$ if $B_{\bf p} = 1$ and in the state $|1\rangle$ if $B_{\bf p} = 0$.  This can be done by carrying out a single-qubit rotation $S$ on the head qubit if and only if the state of the tail qubit is $|0\rangle$ where\cite{modular}
\begin{eqnarray}
S = \frac{1}{\sqrt{1+\phi^2}}\left(\begin{array}{cc} 1 & \phi \\ \phi & -1 \end{array}\right).
\label{smatrix}
\end{eqnarray}

This transformation corresponds to the diagram shown in Fig.~\ref{s}(a) and is defined in terms of the tensor $S^{a}_{bb^\prime}$ which is equal to the matrix $S$ when $a=0$ and for which $S^{1}_{11} = 1$ (the case $S^{1}_{bb^\prime}$ with $b=0$ or $b^\prime=0$ violates the vertex constraint).  A quantum circuit which carries out this transformation (and its inverse since the circuit squares to 1) is shown in Fig.~\ref{s}(b). This circuit can be carried out with 1 CNOT gate and 2 single-qubit rotations.  Like the $F$ circuit, this simple construction is possible because $S^2 = 1$ and $\det S = -1$.   If the two tadpole qubits are initially in the state $|\psi_{B_{\bf p} = 1}\rangle$ the result of carrying out this circuit is the state $|0\rangle |0\rangle$. If the two tadpole qubits are initially in the two-dimensional Hilbert space spanned by the states $\{|\psi_{B_{\bf p}=0},a\rangle, |\psi_{B_{\bf p}=0},b\rangle\}$ then after carrying out this circuit they will be in the space spanned by the states $\{|0\rangle |1 \rangle,|1\rangle |1 \rangle\}$.  In either case the state of the second qubit, i.e. the rotated head of the tadpole, will be equal to $1-B_{\bf p}$.

After carrying out the $S$ circuit on the tadpole, a CNOT gate can be done with the head qubit as the control qubit and a syndrome qubit, initialized to the state $|0\rangle$, as the target qubit.  The syndrome qubit can then be measured and if the result is $0$ then $B_{\bf p} = 1$ for the tadpole (and hence for the original plaquette), and if the result is $1$ then $B_{\bf p} = 0$.

After measuring $B_{\bf p}$ for the tadpole, the final step is to reconstruct the full plaquette.  This can be done by undoing the $S$ circuit on the tadpole and then undoing the $F$-moves.  Putting everything together the resulting measurement circuit for the case of a hexagonal plaquette is the palindromic circuit shown in Fig.~\ref{Bp}.  In this circuit the notation is the same as in the pentagon circuit, with each box corresponding to either the full or reduced $F$ circuit, or the $S$ circuit, and the letters labeling the various ``inputs'' as defined in Figs.~\ref{f},\ref{fprime}, and \ref{s}.   From the structure of the circuit it is clear how this construction generalizes to the case of an arbitrary $n$-sided plaquette.

We again emphasize that the circuit shown in Fig.~\ref{Bp} only measures $B_{\bf p}$ correctly if the vertex constraint $Q_{\bf v} = 1$ is satisfied on each vertex of the initial plaquette at the start of the circuit.  If the vertex constraint is violated on any of these vertices then by definition $B_{\bf p} = 0$ for the plaquette;\cite{levin05} but the circuit will, in some cases, give the wrong result of $B_{\bf p} = 1$.  To see this, consider the action of this circuit on the full $2^{2n}$-dimensional Hilbert space of the $2n$ qubits associated with an $n$-sided plaquette, including those states which violate the vertex constraint. From the structure of the circuit, which performs a unitary transformation on $2n$ qubits and then measures the state of a single qubit to determine $B_{\bf p}$, it is clear that the dimensionalities of the Hilbert spaces for which $B_{\bf p} = 1$ or $B_{\bf p} = 0$ would both be $2^{2n-1}$, i.e. half that of the full Hilbert space. However, once the vertex constraint is taken into account the Hilbert space is greatly reduced.  The dimensionalities of the projected Hilbert spaces for which $Q_{\bf v} = 1$ on each of the $n$ vertices and $B_{\bf p}= 1$ or $B_{\bf p} = 0$ for the plaquette are ${\rm Dim}[B_p = 1] = F_{2n-1}$ and ${\rm Dim}[B_p = 0] = F_{2n+1}$, respectively, where $F_n$ is the $n$th Fibonacci number ($F_{0} = 0$, $F_1 = 1$, $F_2 = 1$, $F_3 = 2$, etc.). For the case of a hexagonal plaquette this means the full $4096 = 2^{12}$ dimensional Hilbert space of twelve qubits is projected down to a space of dimensionality $322 = F_{11}+F_{13} = 89 + 233$ with an 89 dimensional space of states satisfying the plaquette constraint with $B_{\bf p} = 1$.  The reader will be reassured to know we have numerically checked that the circuit shown in Fig.~\ref{Bp} performs the correct measurement of $B_{\bf p}$ on this projected space.

\begin{figure*}[t]
\begin{center}
\includegraphics[scale=.35]{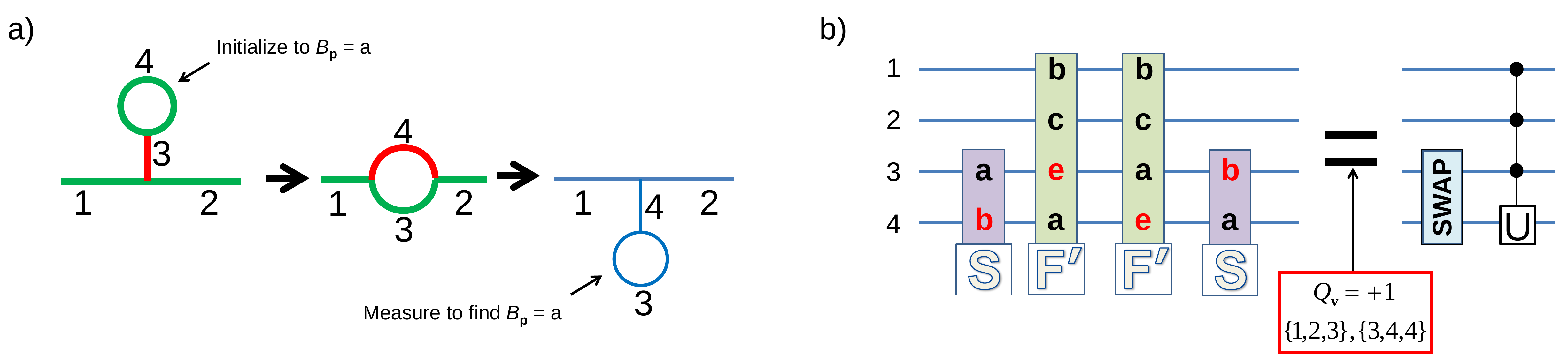}
\end{center}
\caption{(Color online) (a) Sequence of $F$-moves which pulls a tadpole through a line. (b) Four-qubit quantum circuit which initializes a tadpole with an $S$ circuit, carries out the sequence of two $F$-moves shown in (a), and then performs another $S$ circuit so that measuring qubit 3 would yield $B_{\bf p}$ for the new tadpole.  The tadpole is initialized to a state with $B_{\bf p} = 1$ or $0$ depending on whether the initial state of qubit 4 is $|0\rangle$ or $|1\rangle$, respectively.  The circuit equality holds provided $Q_{\bf v} = 1$ on the vertices of the initial lattice.  As in Fig.~\ref{pentagon}(b) these vertices are labeled inside the red box. The $2\times 2$ matrix $U$ is given by Eq.~\ref{umatrix} in the text.} \label{sf_circuit}
\end{figure*}

It should be noted that the requirement that $Q_{\bf v} = 1$ on each vertex before measuring $B_{\bf p}$ may cause problems when extracting error syndromes.  For example, if a faulty measurement of $Q_{\bf v}$ gives 1 for a particular vertex on a plaquette, but the actual value of $Q_{\bf v}$ is 0 for that vertex, then, as described above, the $B_{\bf p}$ measurement circuit for the plaquette will, in some cases, give $B_{\bf p} = 1$ even though the correct value (as it is for any plaquette in which a vertex constraint is violated) is $B_{\bf p} = 0$.  In this paper we have not addressed the important question of whether it is possible to extract error syndromes for the Fibonacci code fault tolerantly, nor the question of precisely how these errors would be corrected.  Our goal has been to construct circuits which, in the absence of errors, can be used to measure $Q_{\bf v}$ and $B_{\bf p}$ in order to begin to get a measure of their complexity.

We can now give our final gate counts for measuring $B_{\bf p}$.  If we choose to reduce all $n$-qubit Toffoli gates to conventional three-qubit Toffoli gates (using $4n - 12$ Toffoli gates, following Ref.~\onlinecite{barenco95_1} as described in Sec.~\ref{Sec:Qv}) then we find that our procedure for an $n$-sided plaquette (with $n\ge 2$) requires $18n - 26$ Toffoli gates,  $8n-5$ CNOT gates and $4n$ single-qubit rotations.  Alternatively, if we consider $n$-qubit Toffoli gates as primitives, then our procedure requires $2n-4$ five-qubit Toffoli gates, 2 four-qubit Toffoli gates, $2n-2$ Toffoli gates, $8n-5$ CNOT gates and $4n$ single-qubit rotations.\cite{2qbgates}  Not surprisingly, this is significantly more demanding than the analogous requirement for the Kitaev surface code, for which only $n$ CNOT gates are needed to measure the plaquette operator for an $n$-sided plaquette. 

Finally we note that there are, of course, many different ways to reduce a given plaquette to a tadpole using $F$-moves, all of which can be used to measure $B_{\bf p}$ and some of which will be more ``parallelizable'' than others.

\section{A Simple Example}

One of the motivations of the present work is to find simple quantum circuits which might feasibly be carried out in the near term and which begin to test some of the key properties of the Fibonacci code.  We have already seen one example of such a circuit, the sequence of five controlled-$F$ gates which results in a two-qubit SWAP gate discussed in Sec.~\ref{Sec:pentagon}.  This circuit is a simplified version of the full seven-qubit pentagon circuit shown in Fig.~\ref{pentagon}(b) and can potentially be used to calibrate the $F$ operation.  In this section we give a similar example --- a four-qubit circuit which first initializes a tadpole into a state with either $B_{\bf p} = 1$ or $0$, and then pulls this tadpole through a line using $F$-moves to produce a new tadpole which can be measured to verify that the value of $B_{\bf p}$ has not changed. As for the pentagon circuit, this four-qubit circuit can be simplified to a two-qubit circuit which, in this case, can be used to calibrate the $S$ operation.

The sequence of operations we consider is illustrated in Fig.~\ref{sf_circuit}(a).  The system consists of a four-edged trivalent lattice and so uses four qubits, labeled 1 through 4 in the figure.  Initially two qubits (1 and 2) are assigned to edges which form a line and the other two qubits (3 and 4) form a tadpole attached to this line.  As always, in what follows we assume that it has been verified that $Q_{\bf v} = 1$ on each of the two vertices of this lattice at the start of the process.  

The first step is to initialize the tadpole in a state with either $B_{\bf p} = 1$ or $0$.  Then, using two $F$-moves, as shown in Fig.~\ref{sf_circuit}(a), the tadpole can be pulled through the line.  The $F$-moves preserve $B_{\bf p}$, and so the intermediate state of this process is a 2-sided plaquette which has been initialized either into the code space if $B_{\bf p} = 1$ or outside of the code space if $B_{\bf p} = 0$.  After the tadpole has been pulled through the line, the two qubits forming the initial tadpole have swapped places --- the head of the tadpole is now the tail and vice versa.  If $B_{\bf p}$ is now measured for the new tadpole the result should yield the same value of $B_{\bf p}$ that the tadpole was initialized to at the start of the process.

The left-hand side of the circuit identity shown in Figure \ref{sf_circuit}(b) is a four-qubit circuit which carries out the procedure described above.  If qubit 4 is initially in the state $|1-a\rangle$ then the first $S$ circuit initializes the tadpole in a state with $B_{\bf p} = a$.  A reduced $F$ circuit then carries out the first $F$-move and produces a 2-sided plaquette with $B_{\bf p} = a$. Next, a second reduced $F$ circuit carries out the second $F$-move producing a new tadpole with $B_p = a$ but with the head and tail of the tadpole interchanged.  Finally, after carrying out an $S$ circuit on this tadpole the state of qubit 3 will be $|1-a\rangle$.  

Note that if qubit 4 is initially in the state $|1\rangle$ so that the tadpole is initialized to a state with $B_p = 0$ then qubit 3 can initially be in either the state $|0\rangle$ or $|1\rangle$ while still satisfying the vertex constraint.  After the first $S$ circuit on the left-hand side of Fig.~\ref{sf_circuit}(b) is carried out the tadpole will then be placed in a quantum superposition of $|\psi_{B_{\bf p} = 0},a\rangle$ and $|\psi_{B_{\bf p}=0},b\rangle$ (See Eqns.~(\ref{bp2}) and (\ref{bp3})).  At the end of the circuit, after being pulled through the line formed by qubits 1 and 2, the tadpole will still be in the two-dimensional $B_{\bf p} = 0$ Hilbert space, but the particular superposition will in general have changed. Direct calculation shows that if qubits 1 and 2 are both in the state $|1\rangle$ then the operation acting on the two-dimensional $B_p =0$ space when pulling the tadpole through the line is given by
\begin{eqnarray}
U = \left(\begin{array}{cc} -\phi^{-2} & \sqrt{1-\phi^{-4}} \\
\sqrt{1-\phi^{-4}} & \phi^{-2}\end{array}\right).
\label{umatrix}
\end{eqnarray}
Otherwise, if either qubit 1 or qubit 2 (or both) are in the state $|0\rangle$ the state of the final tadpole will be the same as that of the initial tadpole with the head and tail qubits swapped.  These cases are all accounted for by the SWAP gate and four-qubit controlled-$U$  operation on the right-hand side of the circuit identity in Fig.~\ref{sf_circuit}(b).  As for the pentagon circuit, this identity only holds when $Q_{\bf v} = 1$ on the two vertices of the initial lattice (again these vertices are labeled inside the red box in the figure).  

\begin{figure}[t]
\begin{center}
\includegraphics[scale=.35]{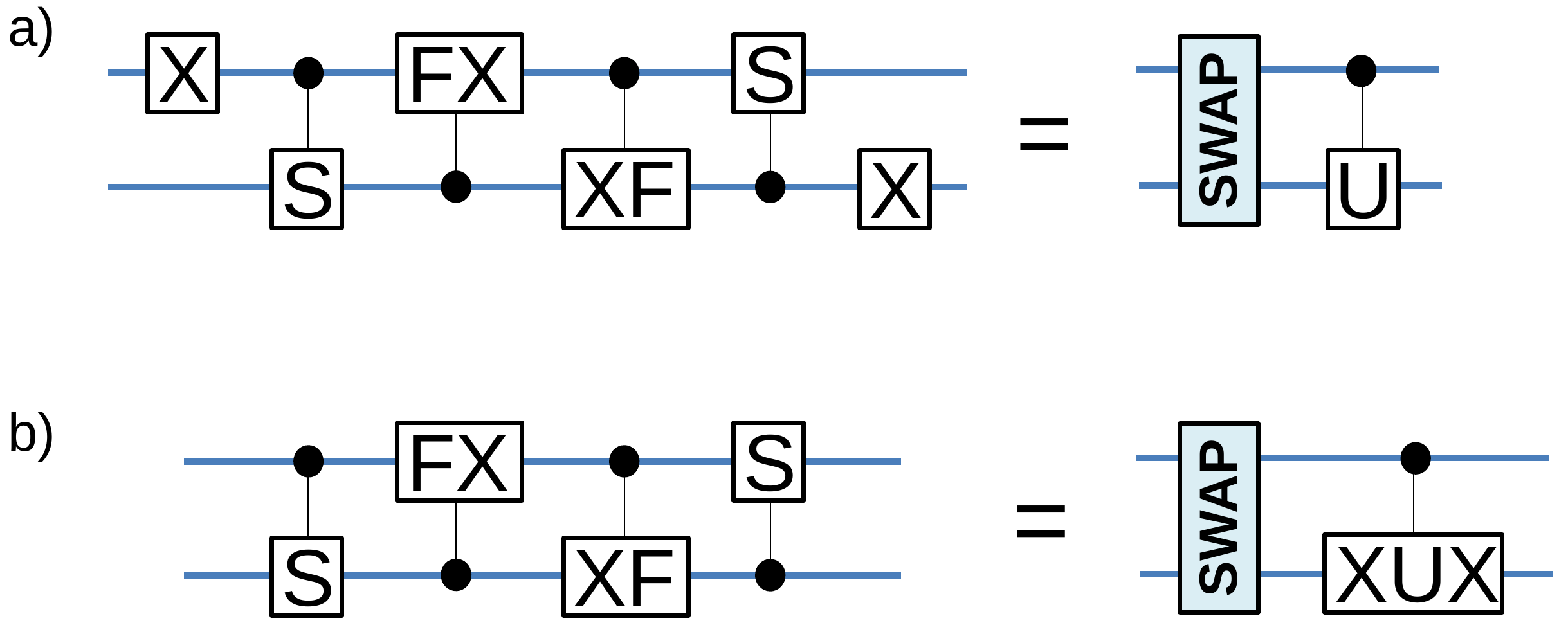}
\end{center}
\caption{(Color online) (a) Simplified two-qubit circuit identity obtained by setting qubits 1 and 2 to the state $|1\rangle$ in the circuit identity shown in Fig.~\ref{sf_circuit}(b).  (b) Equivalent circuit identity obtained by moving the two NOT gates from the left side of the identity shown in (a) to the right side.} \label{sf_simple}
\end{figure}

This four-qubit circuit, which essentially represents initializing a 2-sided plaquette into a state with a given value of $B_{\bf p}$ and then producing a state which can be measured to determine $B_{\bf p}$ after carrying out a different $F$-move than the one used to initialize it, is much simpler than the full circuit for measuring $B_{\bf p}$ for a hexagonal plaquette.  However, it still involves the four-qubit Toffoli gate which appear in the reduced $F$ circuit. As for the pentagon circuit, a simpler two-qubit circuit identity can be found by fixing the states of the qubits which act effectively as control qubits (qubits 1 and 2 in Fig.~\ref{sf_circuit}).  If we fix these qubits to both be in the state $|1\rangle$ we obtain the simplified two-qubit circuit identity shown in Fig.~\ref{sf_simple}(a).  

This circuit identity can be simplified further by multiplying both sides on the left and right by NOT gates which act on the top and bottom qubits, respectively, to obtain the equivalent circuit identity shown in Fig.~\ref{sf_simple}(b).   Note that if the initial state for the circuit shown in Fig.~\ref{sf_simple}(a) is $|1\rangle|0\rangle$, where the first qubit is the top qubit (qubit 3 in Fig.~\ref{sf_circuit}), then the vertex constraint is not satisfied for the full four-qubit circuit.  However, the simplified circuit identity is readily seen to be satisfied in this case.  For all other cases the vertex constraint is satisfied, and so it follows that the expression shown in Fig.~\ref{sf_simple}(a) and the equivalent expression in Fig.~\ref{sf_simple}(b) are exact circuit identities, independent of whether or not the vertex constraint is satisfied.

The key action of the two-qubit circuit on the left-hand side of Fig.~\ref{sf_simple}(b) occurs when the tadpole is initialized in a state with $B_{\bf p} = 1$ for which the tail qubit must start in the state $|0\rangle$.  For this case, after pulling the tadpole through the line the new tadpole must again be in the state with $B_{\bf p} = 1$.  Thus, after accounting for the removal of the two NOT gates, this circuit must take the state $|1\rangle |0\rangle$ to the state $|0\rangle |1\rangle$.  

Like the five controlled-$F$ SWAP circuit in Fig.~\ref{pentagon_simple}, which can be used to calibrate the $F$ matrix, the circuit identity shown in Fig.~\ref{sf_simple}(b) can be used to calibrate the $S$ matrix.  Once $F$ has been fixed by the pentagon circuit, the requirement that the circuit on the left-hand side of Fig.~\ref{sf_simple}(b) takes the state $|1\rangle|0\rangle$ to the state $|0\rangle |1\rangle$ fixes the form of the matrix $S$ (up to an overall phase which is irrelevant for our purposes). Note that in performing this calibration it is not necessary to carry out a full quantum process tomography.  It is sufficient to verify that the circuit identity holds for the initial state $|1\rangle|0\rangle$.  For this case, only the SWAP gate on the right-hand side is relevant since the controlled-$XUX$ gate enters only when the initial state of the second qubit is $|1\rangle$.

\section{Conclusions}

In this paper we have constructed explicit quantum circuits for measuring the vertex and plaquette operators, $Q_{\bf v}$ and $B_{\bf p}$, in the Fibonacci Levin-Wen model.  These operators can be viewed as stabilizers for the Fibonacci code,\cite{koenig10} a surface code for which defects can behave as Fibonacci anyons --- the simplest non-Abelian anyons for which braiding alone is universal for quantum computation.  While the $Q_{\bf v}$ measurement is not significantly more difficult than the analogous measurement for the Kitaev surface codes (for which the defects behave as Abelian anyons), the $B_{\bf p}$ measurement scheme we present here {\it is} significantly more difficult than its Abelian counterpart. While the present scheme is almost certainly not the most efficient one for performing this measurement, given the complexity of the operator $B_{\bf p}$ it is likely that even the most efficient schemes will require a large number of primitive gate operations.  This cost in circuit complexity will then need to be weighed against the gain of not requiring magic state distillation.  The situation is somewhat analogous to comparing the relative merits of performing topological quantum computation with Ising anyons (which requires magic state distillation) to Fibonacci anyons.\cite{baraban10}

It is clear that further work will be needed before such a direct comparison of the resources needed to carry out fault-tolerant quantum computation using the Fibonacci code with that using the Kitaev surface code will be possible.  While recent progress strongly suggests that the Kitaev surface code is the most promising from the practical point of view of trying to build an actual fault-tolerant quantum computer, we believe it is too early to rule out the possibility that the Fibonacci code may have some practical implications.  Even if it does not, we believe the Fibonacci code is of intrinsic interest, in part because computing with it can be viewed as essentially simulating a non-Abelian state of matter on a quantum computer.

Our measurement circuit for $B_{\bf p}$ is built out of circuits which realize $F$-moves and the action of the $S$ matrix on a trivalent lattice.  In addition to our measurement circuits we have also given simpler circuits built out of these $F$ and $S$ circuits. The first is a seven-qubit circuit which can be used to verify that the $F$ circuit satisfies the pentagon equation, as well as a simpler two-qubit circuit which contains the nontrivial content of this equation and fixes the form of the $F$ matrix.  The second is a four-qubit circuit which uses the $S$ circuit to initialize a tadpole (1-sided plaquette) in a state with either $B_{\bf p} = 1$ or $0$, carries out a sequence of $F$-moves to pull the tadpole through a line, and then produces a state which can be measured to determine $B_{\bf p}$ for the new tadpole.  For this circuit we have also given a simpler two-qubit circuit which, once $F$ has been fixed by the pentagon circuit, fixes the form of the $S$ matrix.  These simple two-qubit circuits (Figs.~\ref{pentagon_simple} and \ref{sf_simple}) may be useful for calibrating the $F$ and $S$ operations.

A recurring theme in this work has been the need for $n$-qubit Toffoli gates (with $n =3,4$ and 5) when computing with the Fibonacci code. These $n$-qubit Toffoli gates arise as a natural consequence of the non-Abelian nature of this code. We believe the possibility of using non-Abelian surface codes such as the Fibonacci code for fault-tolerant quantum computation provides further motivation for developing experimental techniques to directly carry out accurate $n$-qubit Toffoli-class gates.

\ 

\acknowledgments NEB acknowledges support from US DOE Grant No. DE-FG02-97ER45639 and DDV is grateful for support from the Alexander von Humboldt foundation.


\begin{thebibliography}{}
\bibitem{modular} $S$ is the so-called modular $S$ matrix for Fibonacci anyons.
\end{thebibliography}
\end{document}